# Electron Counting Capacitance Standard with an improved five-junction *R*-pump

Benedetta Camarota[1], Hansjörg Scherer[1], Mark W Keller[2], Sergey V Lotkhov[1], Gerd-Dietmar Willenberg[1], and Franz Josef Ahlers[1]

[1]Physikalisch-Technische Bundesanstalt (PTB) Braunschweig, Bundesallee 100, 38116 Braunschweig, Germany

[2]National Institute of Standards and Technology (NIST), 325 Broadway, Boulder, CO 80305-3328, USA

**Abstract**

The Electron Counting Capacitance Standard currently pursued at PTB aims to close the Quantum Metrological Triangle with a final precision of a few parts in $10^7$. This paper reports the considerable progress recently achieved with a new generation of single-electron tunnelling devices. A five-junction *R*-pump was operated with a relative charge transfer error of five electrons in $10^7$, and was used to successfully perform single-electron charging of a cryogenic capacitor. The preliminary result for the single-electron charge quantum has an uncertainty of less than two parts in $10^6$ and is consistent with the value of the elementary charge.

Single-electron tunneling, 73.23.Hk

Charge measurement, 84.37.+q

Coulomb blockade, 73.23.Hk

Electric charge, 41.20.Cv, 84.37.+q

Nanoelectronic devices, 85.35.-p

Tunnel junction devices, 85.30.Mn

1. **Introduction**

The principle of the Electron Counting Capacitance Standard (ECCS) experiment – pioneered [1] and first demonstrated [2] by NIST - is to charge a capacitor with a known number of electrons and to measure the resulting voltage across the capacitor electrodes. When the voltage is measured in terms of the Josephson effect and the capacitive impedance is measured in terms of the quantum Hall effect (QHE), this experiment is one of the versions of the Quantum Metrological Triangle (QMT) [3][4][5].

QMT experiments test the fundamental consistency of three quantum electrical effects involving the elementary charge $e$ and the Plank constant $h$, expressed in the following theoretically predicted relations: the Josephson frequency-to-voltage conversion factor being $K_J = 2e/h$, the quantum Hall impedance being $R_K = h/e^2$, and the single-electron tunnelling charge quantum being $Q_S = e$. Both theory and experiments support the assumption that these three relations are exact, but efforts continue to determine upper bounds on possible corrections in actual devices that embody the quantum effects. The importance of such efforts is confirmed by recent CIPM recommendations [6]. Currently, possible corrections to these relations are believed to be smaller than one part in $10^6$ [5]. A QMT experiment performed with a relative uncertainty of a few parts in $10^7$ or better would strengthen the confidence in the understanding of the quantum electrical effects, and may also contribute to the determination of corresponding phenomenological constants, in particular $K_J$ [5][7].

Charging the capacitor in the ECCS is performed using a single-electron tunnelling (SET) pump to transfer charges one-by-one and an SET transistor to monitor the charging process. In the PTB setup [8][9], the SET pump contains on-chip resistors in series with the junctions, a configuration known as an *R*-pump. The cryogenic capacitor (CryoCap) used in the ECCS is located near the SET devices, whose operation requires temperatures below 100 mK, and has a low loss due to the vacuum gap between the electrodes [10]. The PTB version of the CryoCap, with coaxial stainless steel electrodes and capacitance within $10^{-4}$ pF of $C_{cryo}$ = 1 pF, is described in [9] and [11]. Since the publication of [9], substantial improvements have been made to the ECCS experiment at PTB. In particular, changes to the layout of the five-junction *R*-pump have reduced the single-electron transfer error by about one order of magnitude and have also made operation of the pump more robust and repeatable. This enabled the first single-electron charging of the CryoCap at PTB and the preliminary result for the QMT reported in this paper . In addition, the CryoCap impedance was traced to the von-Klitzing constant by using high-precision impedance measurement bridge techniques and the ac QHE [12]. This second advance, combined with ongoing work on charging the CryoCap, should enable the PTB version of the ECCS to close the QMT with a precision better than the nine parts in $10^7$ achieved by NIST [2][13].

2. **Single-electron tunnelling devices: layout and performance**

The SET devices used in the experiment reported here consist of a five-junction single-electron *R*-pump and an SET transistor, all based on Al-AlO$_x$-Al tunnel junctions and fabricated on a fused quartz substrate (see Figure 1 for the schematic circuit). The single-electron pump is equipped with on-chip chromium micro-strip resistors, each with resistance $R_{Cr} \approx$ 50-60 k$\Omega$. The resulting modification of the effective electromagnetic environment seen by the junctions has been shown to suppress unwanted co-tunnelling events [14] that are presumed to compromise the pumping accuracy [9][14].

Between the *R*-pump and the electrometer there is a metallic region (pad island) serving as the landing pad for a cryogenic needle switch. The switch, which operates like an electromagnetic relay, provides the connection to the CryoCap located several centimeters from the chip. Some of the preliminary characterization measurements on the *R*-pump-and the SET transistor, such as current-voltage curves, are performed-with the switch closed [9][15][16].

The transistor is capacitively connected to the pad island via an interdigitated gate with capacitance $C_{ID} \approx 1$ fF, and serves as an electrometer to detect the charge state - or the potential - of the pad island. When single-electron resolution of the pad island charge state is required (for instance, when tuning the pump's working point and determining its transfer error rate as described below), the switch is kept open [2][9][15]. In this configuration, high-fidelity charge detection with single-electron resolution is achieved when the parasitic stray capacitance $C_0$ of the pad island to ground is minimized [15]. Using a low dielectric constant substrate (fused quartz, giving $C_0 \approx 20$ fF) resulted in a charge divider ratio of about 1:20 at the electrometer input, allowing reliable single-electron detection [9]. A detailed description of the device fabrication, as well as the realization of on-chip silicon shunts to protect the device against electrostatic breakdown, is provided in [9][17].

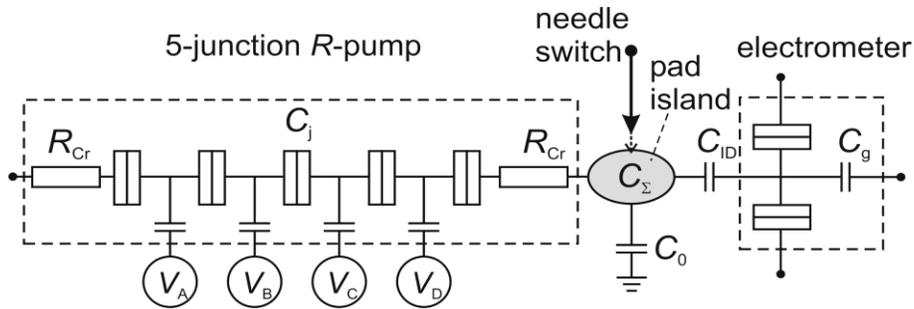

Figure 1: Schematic of the SET circuit. The chip contains an *R*-pump with five nanometer-scale metallic tunnel junctions of capacitance $C_j$, two in-series chromium resistors, and an SET electrometer. The metallic pad island connecting pump and electrometer can be contacted electrically to one electrode of the cryogenic capacitor (not shown) by closing the cryogenic needle switch. By adjusting the dc voltages $V_A \ldots V_D$ on the four gates of the *R*-pump, the device is tuned to its working point.

In the previous generation of devices, the layout of the SET components was as illustrated in the upper panel of Figure 2. One such device, presenting the best performance for a five-junction single-electron *R*-pump thus far, showed a single-electron transfer error corresponding to a few parts in $10^6$ [9]. More devices with the same layout from different fabrication batches were subsequently tested, but all showed worse performance. Furthermore, the adjustment of the dc gate voltages for minimum single-electron transfer error was difficult for all the investigated devices because their performances were irreproducible [18].

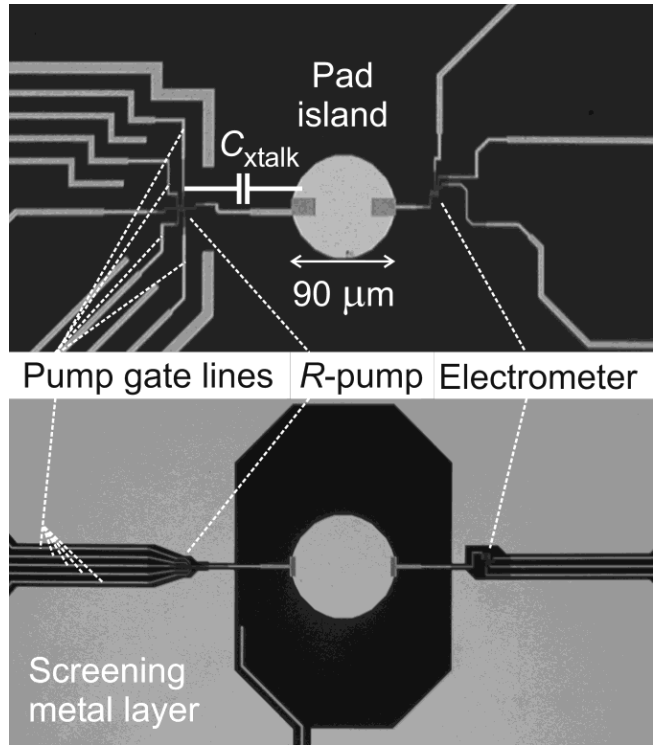

Figure 2: Micrographs of the SET devices, comparing the chip layout of the old (upper panel) and new (lower panel) design. In the old design, significant cross-talk capacitances between the gate lines and the pad island (indicated by $C_{xtalk}$ for gate C, cf. Figure 1) occurred due to the mostly unscreened geometry. In the new layout, the pad island is better screened from the SET devices, and the geometric routing of the pump gate lines is more favourable.

Further experimental investigations revealed considerable parasitic cross-capacitance $C_{xtalk}$ between the pump gate lines and the adjacent metallic pad island due to the sample layout geometry (Figure 3) [18]. Applying a known voltage to a pump gate caused a change in the electrometer signal that could be converted into a charge on the pad island (by capacitive division using the values provided above), and dividing the induced charge by the applied voltage gave the effective $C_{xtalk}$ for that pump gate. $C_{xtalk}$ was measured for each of the four pump gates on multiple devices with the same layout, and Figure 3 shows the results for three such devices. For the largest values of $C_{xtalk} = 0.8$ fF, the typical pulse amplitude of 6 mV applied to the gates during pumping would induce a charge of 30 electrons on the pad island. This corresponds to a change of about 250 µV in the voltage drop across the pump device, which is expected to significantly increase single-electron transfer errors, according to the data reported by NIST on 7-junction single-electron pumps [19].

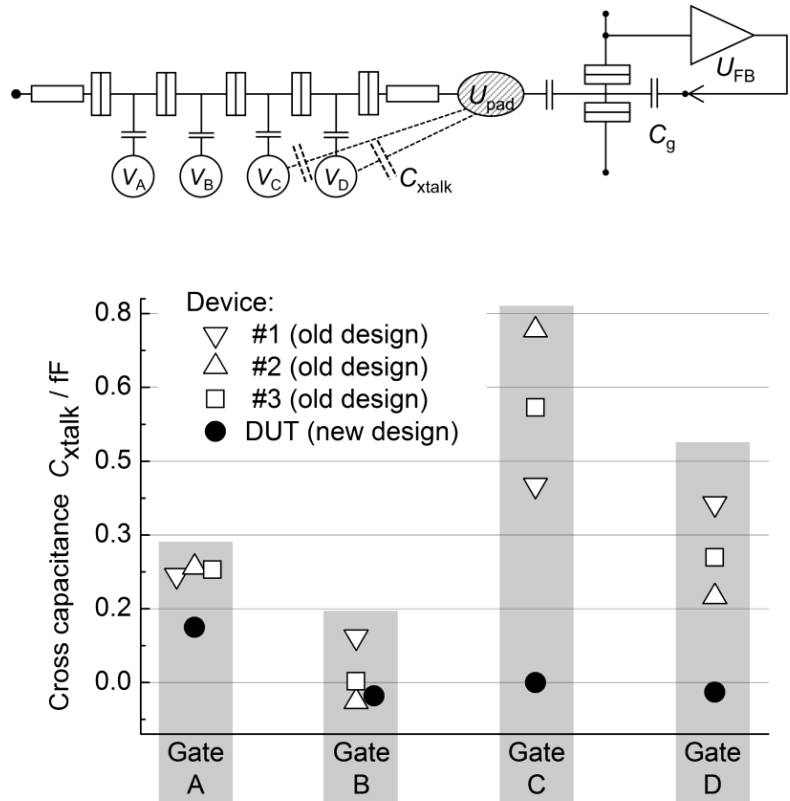

Figure 3: Schematic of the SET device and the feedback circuit (upper panel). Cross-talk capacitances between the pump gates C and D and the pad island are outlined for clarity. The lower panel shows effective cross-talk capacitance values $C_{xtalk}$ for the four gates of three devices with the old layout (white triangles and squares) compared to $C_{xtalk}$ derived for the new layout (black dots). The slightly negative $C_{xtalk}$ values observed for gates B and D are an artefact. They are caused by a cross-capacitance cancellation circuit [16][20], which was here apparently slightly mistuned, thus over-compensating the undesired, inevitable cross-polarization from the pump gates to the neighbouring pump islands.

In order to minimize $C_{xtalk}$, an improved lithographic device layout was implemented (Figure 2). By modifying the routing of the $R$-pump gate lines and by introducing a metallic shielding plane, it was possible to significantly reduce $C_{xtalk}$ in comparison with the former layout, as investigations on the Device Under Test (DUT) showed (Figure 3). Further measurements on the DUT confirmed a definitive improvement in the pumping performance and reproducibility, as presented in the following section.

### 3. Improvement of single-electron pump performance

The experiments were carried out in a dilution refrigerator system at its base temperature of 30 mK [8][9]. The preliminary tuning procedure and the operation of the pump device were performed using the methods and instrumentation (gate voltage pulse generator, cross-capacitance cancellation circuit, and the voltage feedback circuit used with the SET electrometer) originally developed by NIST and described in detail in [21][16].

The proper single-electron transfer functioning of the $R$-pump was verified through "shuttle pumping" tests: the pump electronics were set to pump one electron in and out from the pad island, with a chosen wait time $t_{wait}$ between successive pump events, while the electrometer was used to monitor the charge state of the pad island via the feedback circuitry (see Figure 3). When the wait time was set long enough (> 0.1 s), the electrometer detection bandwidth could resolve the pumping of each electron, as shown for $t_{wait}$ = 0.4 s in Figure 4. The electrometer signal $U_{FB}$ clearly follows the single-charge shuttling events, hence proving the correct function of the pump and the electrometer.

The Relative Transfer Error (RTE) for single electrons was determined in a shuttle pumping measurement following the method described in [15][21][16]. The pump was operated at the shuttling frequency of 0.5 MHz (using a clock frequency of 4 MHz, and setting $t_{wait}$ = 0 μs, the same settings later employed for the capacitor charging phase of the ECCS experiment), and the feedback signal $U_{FB}$ at the electrometer gate $C_g$ was recorded. Figure 5 shows a result corresponding to an RTE of a few parts in $10^7$, which was typical for the DUT. In comparison with results on the old layout devices [9], this RTE performance represents an improvement of about one order of magnitude. Furthermore, the RTE values were stable with time once the device had been cold for about one week, as shown for instance in Figure 5 b. Adjustments to the working point were then required about once a day, to account for shifts in background charges near the pump islands. Also, optimum RTE values were repeatable after thermal cycling between base and room temperature.

More generally, the pumping accuracy of this five-junction $R$-pump is at present exceeded only by that reported by NIST on a seven-junction pump, showing lower RTE values by about one order of magnitude [15].

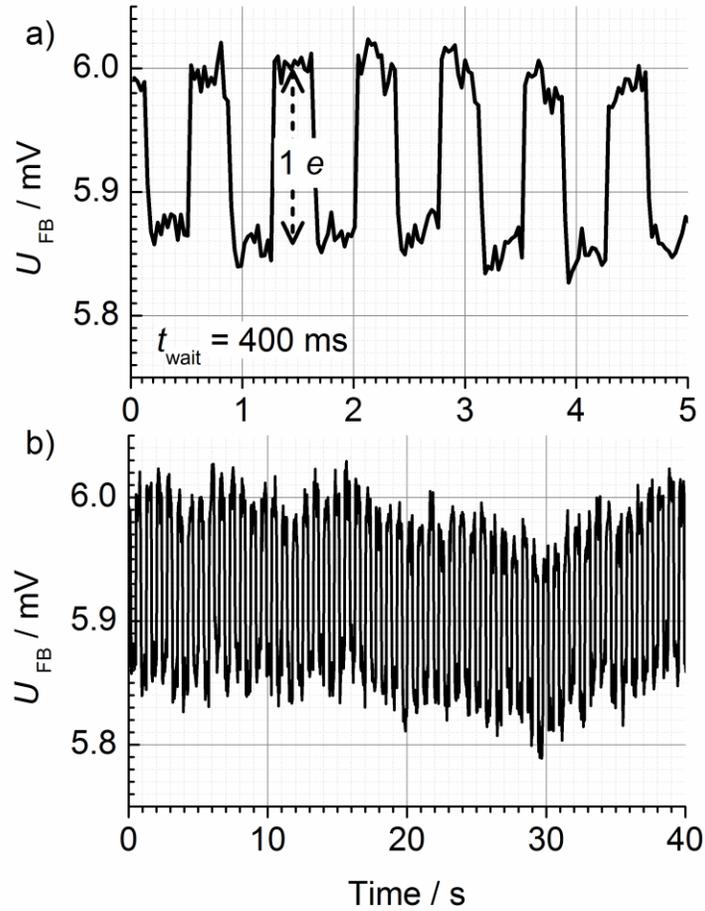

Figure 4: Single-electron shuttle pumping. The *R*-pump was operated in shuttle mode with a wait time $t_{wait} = 0.4$ s between the transfer events (a). One electron on the pad island corresponds to $\Delta U_{FB} \approx 0.14$ mV. SET shuttling was maintained without errors over more than 100 cycles (b). The discernible signal drift is caused by background charge shifts in the vicinity of the electrometer island.

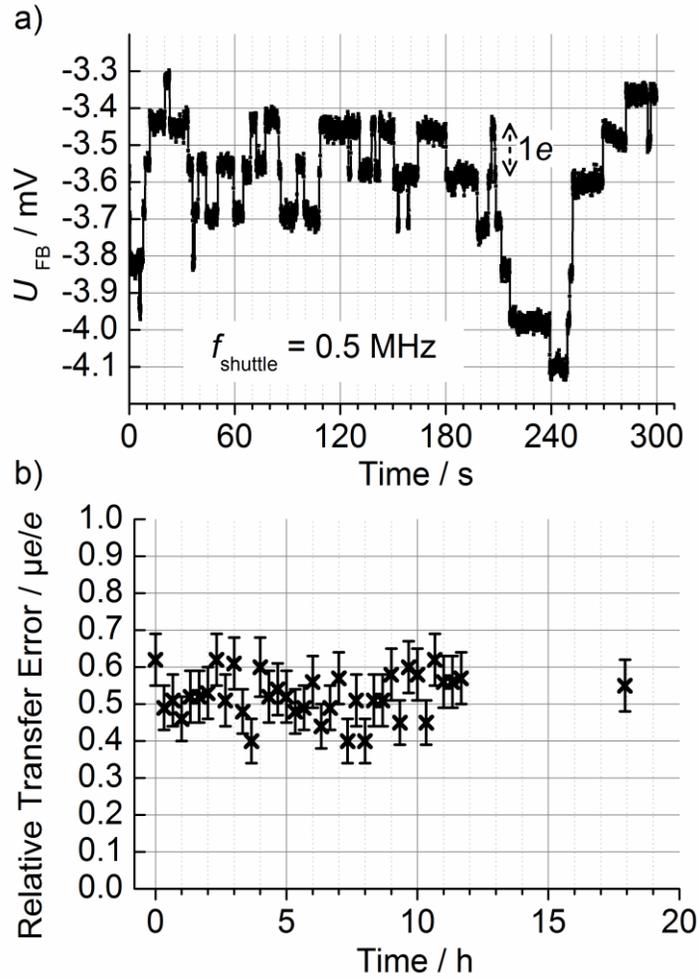

Figure 5: (a) Typical single-electron transfer error measurement at 0.5 MHz shuttling frequency. The electrometer, bandwidth-limited to about 1 kHz, records only the error events. During 300 s of monitoring, 56 single-electron transfer error events occurred, each indicated by the steps in the graph. The corresponding error rate of 0.2 s$^{-1}$, related to the shuttling frequency, results in a relative transfer error RTE ≈ 0.4 µe/e. (b) Repeated measurements of the RTE over a time span of 18 hours without re-tuning of the pump. The RTE varied within about 2×10$^{-7}$ (typical result).

Once the pumping was stopped, the electrometer detected spontaneous single-electron charge fluctuations on the pad island caused by unwanted random tunnelling events. Those events can be triggered by thermal activation, background charge activity, or electromagnetic interference[1] in the system [16][19][22]. For the device under investigation, typical hold times - i.e. the average time between such events - ranged between $t_{hold}$ = 20 s and 30 s. In the best case $t_{hold}$ = 67 s was found (Figure 6).

---

[1] The effectiveness of the rf attenuation components in the lines of the setup was tested by measuring the saturation of the current vs gate voltage modulation of an SET transistor device around 40 mK when the temperature of the dilution refrigerator was varied. The result of this measurement indicated good thermalization of the on-chip SET devices.

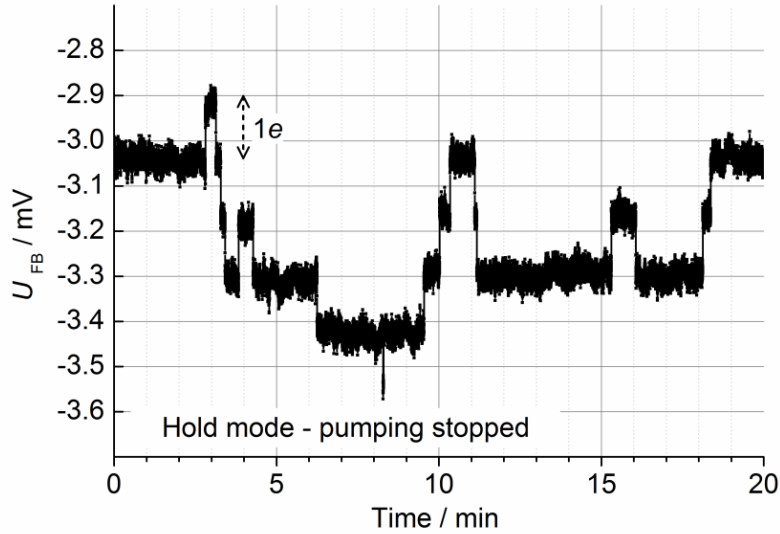

Figure 6: Hold mode measurement over 20 min. The average time between SET events was $t_{hold}$ = 67 s.

As the DUT satisfied the following basic requirements, namely (i) high-fidelity performance of the single-electron transfer, (ii) hold times long enough for voltage measurements on the CryoCap in between charging cycles, and (iii) stability of the working point over several hours, it was suitable to be used for the capacitor charging phase of the ECCS experiment, illustrated in the next section.

4. **Capacitor charging by single electrons**

The setup for the capacitor charging phase of the ECCS experiment, described in [2][13], is schematically shown in Figure 7. With the needle switch NS1 closed, charge was transferred from the SET pump to one side of the CryoCap (the "low potential" electrode). To minimize transfer errors during the charging of the capacitor [19], the voltage across the pump was kept near zero by using the electrometer as a null detector for driving a feedback circuit, applying a compensating voltage to the "high potential" electrode of the CryoCap. This was also necessary to ensure that all transferred charge was collected on the capacitor electrodes, and not on the stray capacitances (not shown in Figure 7) between the pad island and ground. The feedback voltage, i.e., the voltage $U_{cryo}$ across the CryoCap electrodes, was measured by using a high-resolution voltmeter (Agilent model 3458A[2]), calibrated with a Josephson voltage standard.

---

[2] This commercial instrument is identified in this paper to adequately specify the experimental procedure. Such identification does not imply any endorsement or recommendation by PTB or NIST, nor is it intended to imply that the equipment identified is necessarily the best available for the purpose.

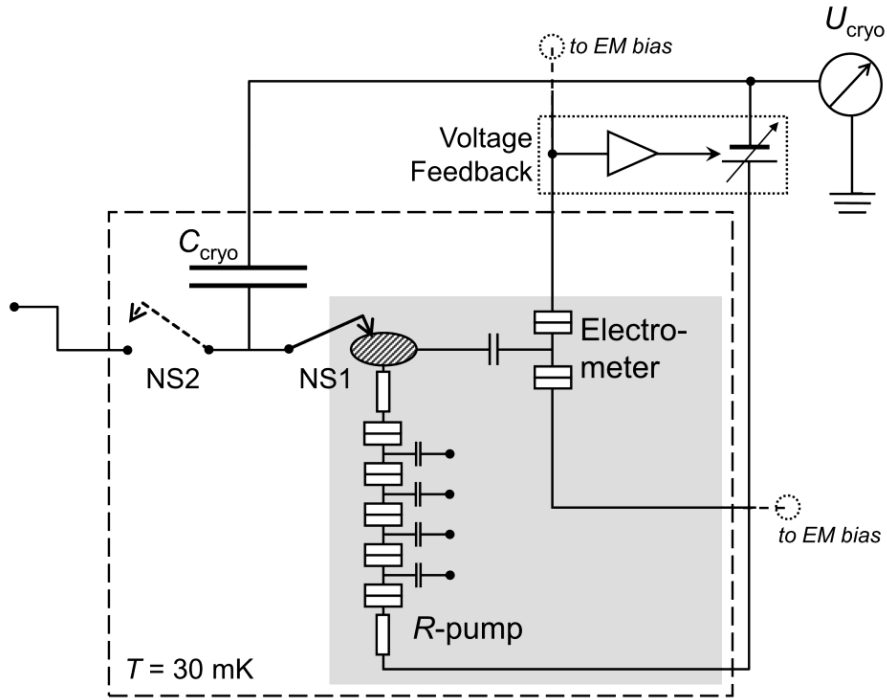

Figure 7: Schematic circuit for the capacitor charging phase of the ECCS experiment. The needle switch NS1 is closed to connect the *R*-pump to the CryoCap via the pad island (SET chip shown shaded). The second needle switch NS2 is opened in the capacitor charging phase and closed when connecting the CryoCap to the capacitance bridge for measuring $C_{cryo}$. The bias circuit for the electrometer source/drain terminals (dotted line ends in the figure) is not shown here for clarity.

After tuning the *R*-pump for its optimum working point (i.e., adjusting the dc voltage settings on the pump gates for minimum transfer errors), and determining its RTE and hold time, the device was connected according to the circuitry described in Figure 7. Then, the capacitor charging phase was initiated as follows [2][13][16]. The pump electronics were set for alternating transfer of about $\Delta N = 3 \times 10^7$ electrons to charge the 1 pF capacitance between ± 2.5 V (see Figure 8). The feedback voltage $U_{cryo}$ was monitored (Figure 7) during the charging-discharging ramps, as well as during the plateaux between ramps when the pump was stopped for about 15 s. Four sequences of several continuous charging cycles (each one called a "run" and lasting between 10 min and 40 min) were performed in the same day (Figure 8).

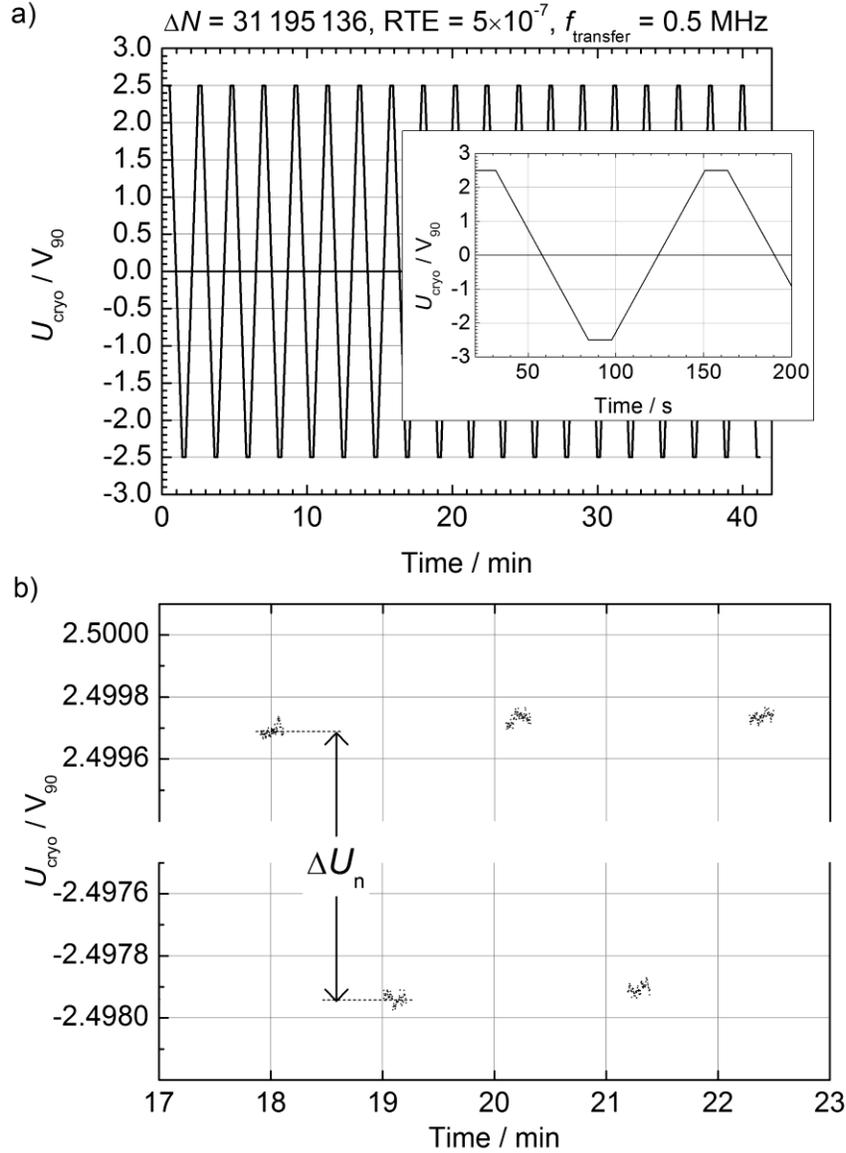

Figure 8: Cyclic charging of the CryoCap to ± 2.5 V. In each cycle the *R*-pump transferred 31 195 136 electrons at clock frequency of 0.5 MHz. The RTE value was determined before this measurement to be five parts in $10^7$, corresponding to ± 16 electrons for the nominally pumped number of electrons in each cycle. Between each cycle the pump was stopped for about 15 s wait time for the measurement of the feedback voltage $U_{cryo}$. (a) Complete data set for a 40 min. run, with time axis expanded in inset. (b) Expanded view of a few plateaux, with $\Delta U_n$ being the voltage difference between the two adjacent plateaux in the *n*th charging cycle.

Also, a preliminary estimate of leakage resistance between the CryoCap electrodes was derived following the rationale described in [2][13]: the capacitor was charged to a voltage of 10 V with the *R*-pump, then charging was stopped and the feedback voltage $U_{cryo}$ was monitored for 30 minutes. From the drift of $U_{cryo}$, a lower bound of $R_{leak} \approx 1 \times 10^{21}$ Ω for the leakage resistance was derived. However, since the measurement uncertainty was limited by the intrinsic drift of the SET electrometer, the real leakage resistance of the CryoCap may be considerably higher.

After the CryoCap charging runs, and within the same cooling cycle, needle switch NS1 was opened and NS2 was closed (Figure 7) to allow for a measurement of $C_{cryo}$ using a commercial capacitance bridge (Andeen-Hagerling model 2500A[2]). The result was $C_{cryo}$ = 1.000 070 (1) pF with an uncertainty of less than one part in $10^6$, limited by the specified uncertainty of the capacitance bridge.

## 5. Results

The preliminary result of this paper will be presented in terms of an SI value for the SET charge quantum obtained from

$$Q_S = C_{cryo} \times \Delta U / \Delta N, \tag{1}$$

where for each run of cyclic charging $\Delta U$ is the mean of the $\Delta U_n$ values (see Fig. 8b). While the capacitance bridge measures $C_{cryo}$ in terms of the SI farad, the voltmeter is calibrated using the value of $K_J$ adopted in 1990, $K_{J-90} \equiv 483\,597.9$ GHz/V [23], and thus measures $U_{cryo}$ in terms of the "1990 volt" $V_{90}$. The conversion to SI volts (V) is straightforward: the notation for the quantity $U$ as the product of its numerical value and its unit, $U = \{U\}_{SI}\,V = \{U\}_{90}\,V_{90}$, combined with the relation between $V_{90}$ and the SI volt (V), i.e. $V_{90} / V \equiv K_{J-90} / K_J$, gives

$$Q_S = \frac{C_{cryo} \times K_{J-90} \times \{\Delta U\}_{90}\,V}{K_J \times \Delta N}$$

for equation (1).

In this paper the latest 2010 CODATA value, $K_J = 483\,597.870$ GHz/V, having a relative standard uncertainty of 2.2 parts in $10^8$ [24], was used.

For each run, following the approach described in section 4.1 of [13], $\Delta U$ is assigned a statistical (Type A) uncertainty that combines its standard deviation, derived from the scatter of the $\Delta U_n$ values, with a contribution from the asymmetry between up and down ramps. Table 1 lists these uncertainties, as well as several other uncertainties from the preliminary analysis. The largest systematic (Type B) uncertainty is that of the capacitance bridge used to measure $C_{cryo}$. Other systematic uncertainties, as described in [13], have been found through a preliminary analysis to be significantly lower than one part in $10^6$ and will be presented in a future paper.

|   | Component | Relative uncertainty in µC/C | Comment |
|---|---|---|---|
|   | *Pumping phase* |   |   |
|   | Type A |   | *see Fig. 9* |
| a | Run 1 | 3.60 |   |
| b | Run 2 | 3.83 |   |
| c | Run 3 | 2.49 |   |
| d | Run 4 | 3.82 |   |
| e | $U_{\text{cryo}}$ traceability to $K_J$ | 0.5 | *measurement with DVM (Type B)* |
| f | Pump error (RTE) | 0.5 | *typical value for this DUT (Type B)* |
|   | *Bridge phase* |   |   |
| g | Type A | 0.04 |   |
| h | $C_{\text{cryo}}$ traceability to $R_K$ | 0.87 | *bridge manufacturer specs (Type B)* |
| i | Loading corrections | 0.4 | *estimate according to [13] (Type B)* |
| k | Other small contributions | 0.5 | *(Type B)* |
| l | Type B only | 1.29 | *root-sum-square of lines e, f, h, i, k* |
|   | **Total** |   | *root-sum-square of lines g, l, and line* |
| m | Run 1 | 3.82 | *a* |
| n | Run 2 | 4.04 | *b* |
| o | Run 3 | 2.80 | *c* |
| p | Run 4 | 4.03 | *d* |
| q | Relative SDOM | 1.05 | *from the scatter of the data points (Type A) of ECCS runs 1-4 (Fig. 9)* |
| r | **Final relative uncertainty** | **1.66** | *root-sum-square of lines l and q (Fig. 9, rightmost point)* |

Table 1: Preliminary uncertainty budget for the ECCS with conservatively estimated values for the main uncertainty contributions with $u_{\text{rel}} > 0.3 \times 10^{-6}$. Further Type B contributions, described in the text, summarized in line *k* of this table, are each smaller than $0.3 \times 10^{-6}$.

Figure 9 shows the values of $Q_S$ for each of the four ECCS runs. The mean of these four values (shown as the rightmost data point in Figure 9) is

$$Q_S = 1.602\,176\,1\,(27) \times 10^{-19}\ \text{C} \qquad [u_{\text{rel}} = 1.66 \times 10^{-6}], \qquad (2)$$

where the number in parentheses is the standard uncertainty referred to the last digits of the quoted value and the number in square brackets is the relative standard uncertainty. This value agrees with the most recent CODATA recommended SI value of the elementary charge [24],

$$e = 1.602\,176\,565(35) \times 10^{-19}\ \text{C} \qquad [u_{\text{rel}} = 2.2 \times 10^{-8}], \qquad (3)$$

which is shown by the dashed line in Figure 9.

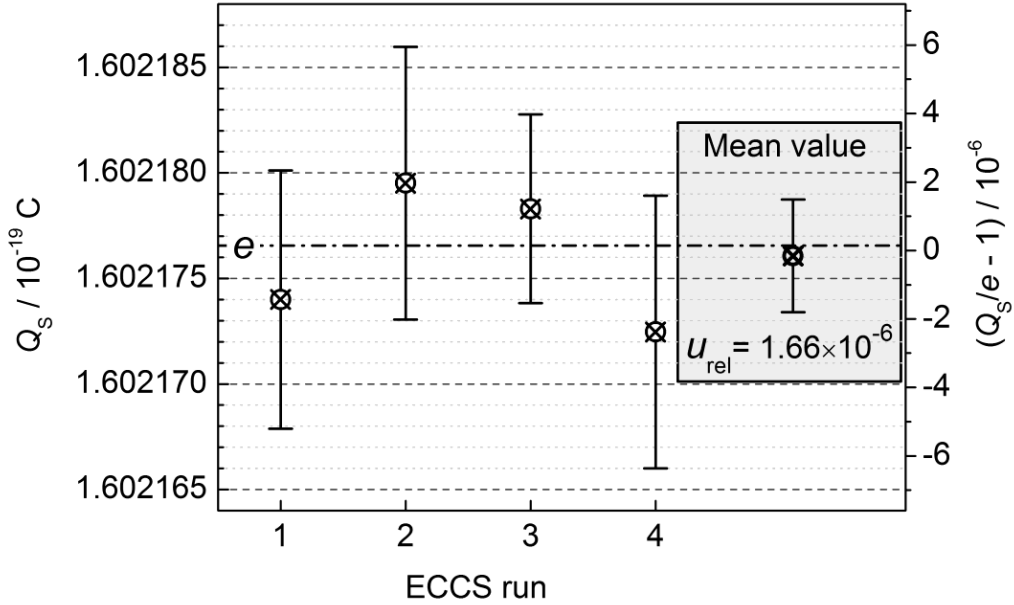

Figure 9: $Q_s$ as determined from four runs of the ECCS experiment, and the final result, given by their mean value. The dashed horizontal line corresponds to the CODATA value for the elementary charge $e$ [24]. The uncertainty bars shown correspond to the standard uncertainty for each data point (coverage factor $k = 1$). The deviation of the determined mean $Q_s$ value from $e$ is three parts in $10^7$.

## 6. Conclusion and Outlook

The advances described here have enabled repeatable and robust charging of a cryogenic capacitor with a known number of electrons for the first time at PTB. While the main result in equation (2) is preliminary because some Type B uncertainties have not been determined, none of these is expected to be as large as the Type A uncertainty of the result presented here. It is thus possible to conclude that this result closes the QMT with a relative uncertainty of $1.66 \times 10^{-6}$.

Expressing the result in equation (2) as a ratio between $Q_S$ and $e$ gives

$$Q_S/e - 1 = (-0.31 \pm 1.66) \times 10^{-6}. \qquad (4)$$

As discussed in detail in [5], a QMT experiment performed with an uncertainty of about one part in $10^6$ or above is primarily testing the SET "leg" of the QMT, since there is confidence about the quantum Hall and Josephson legs in this uncertainty regime.

The experimental conditions achieved to date are not completely optimized, and some aspects of the PTB design have not yet been fully exploited. Further improvements are expected in the following areas:

(i) Optimization of the operational parameters for the five-junction $R$-pump as well as the testing of other devices having the new layout is currently in progress. A relative transfer error of about one part in $10^7$ should be possible.

(ii) The Type A uncertainty of the final result is expected to improve with a larger number of charging cycles, and by charging the CryoCap to higher voltages, as charging up to $\pm 10$ V has already been demonstrated during a test run. Further improvement will result from a more precise preparation of the feedback circuit before charging, and possibly from devices with lower SET electrometer noise. A final Type A uncertainty of about one part in $10^7$ should be possible.

(iii) Dedicated equipment for high-precision voltage measurements based on the Josephson voltage standard with an expected uncertainty below one part in $10^8$ is available at PTB [25] and will be exploited in future runs of the ECCS experiment.

(iv) A direct link between the CryoCap and the ac QHE, using a high-precision impedance bridge technique developed at PTB, has been demonstrated [12]. This will allow $C_{\text{cryo}}$ to be measured in terms of $R_K$ with a relative uncertainty of three parts in $10^8$ [26]. As a result, the term that dominated the uncertainty of the NIST result [13] will be negligible in the PTB experiment.

(v) The PTB CryoCap benefits from a very small frequency dependence between about 10 mHz (the effective charging frequency in the ECCS) and 1 kHz (the operating frequency of the capacitance bridge). A conservative estimate based on [27] shows that the larger distance between the capacitor electrodes (5 mm for the PTB design *vs*. 50 μm for the NIST design) makes the uncertainty due to this frequency dependence smaller than two parts in $10^8$ [9].

With these improvements, a total uncertainty of a few parts in $10^7$ can be expected for the ECCS at PTB. At this level, the experiment will give new information about possible corrections to both the SET charge quantum $Q_S$ and the Josephson constant $K_J$. In particular, it would offer a new observational equation for the CODATA analysis of possible corrections to these effects.

## 7. Acknowledgements

The Electron Counting Capacitance Standard (ECCS) experiment has been pursued at PTB within the EURAMET joint research project "REUNIAM" (March 2008- May 2011). The research conducted within this project has received funding from the European Community's Seventh Framework Programme, ERA-NET Plus, under Grant Agreement No. 217257.

The authors gratefully acknowledge the following contributions:
Jürgen Schurr and Veit Bürkel for support with the capacitance measurements, and Ralf Behr for voltmeter calibration,
Ulrich Becker, Michael Busse, Gerd Muchow, and Norbert Tauscher for providing valuable technical assistance,
and Blaise Jeanneret (METAS) for lending spare electronic equipment.